\documentclass[twoside,12pt]{article}
\usepackage[utf8]{inputenc}
\usepackage{authblk}

\usepackage{graphicx}
\usepackage{a4wide}
\usepackage{palatino}
\usepackage{amsmath}
\usepackage{amssymb}
\usepackage{isotope}
\usepackage[shortlabels]{enumitem}
\numberwithin{equation}{section}
\usepackage{hyperref}

\def\vec#1{\boldsymbol{#1}}

\counterwithout{figure}{section}

\begin{document}

\title{Stability of the Hydrogen Molecule\\ and Related Issues%
\footnote{Contribution to the special issue of \textit{Pure and Applied Functional Analysis}, on the occasion of J\"urg Fr\"ohlich's 80th birthday, edited by Volker Bach, Simeon Reich and Alexander Zaslavski}}

\author{Jean-Marc Richard}
	\affil{Universit{\'e} Lyon 1, CNRS, IP2I Lyon, UMR 5822 \authorcr
	{\small\sl	4 rue Enrico Fermi, 69622 Villeurbanne, France }}
\date{\small \today}
\maketitle

\begin{abstract}
We review the collaboration that led to the first rigorous proof of the stability of the hydrogen molecule within quantum mechanics and discuss several related issues concerning few-charge systems. Particular emphasis is placed on the role of symmetry breaking, the stability domains of Coulombic few-body systems, and some applications to exotic hadrons in the quark model.
\end{abstract}

\vskip .2cm
\centerline{AMS subject classifications: 81-03, 81Vxx\hfill}
\centerline{Keywords: hydrogen molecule, positronium molecule, multiquark hadrons\hfill}

\section{History}

In the early 1990s, Andr\'e Martin, who was then working at CERN, realized that J\"urg and I had been studying independently the stability of the hydrogen molecule, H$_2$, using rather different approaches. Acting as a friendly intermediary, he put us in contact, and we decided to combine our efforts. The outcome was the first rigorous proof of the stability of the hydrogen molecule.

I received several cordial and somewhat ironical remarks from experimental colleagues, along the lines of: ``It is reassuring to learn, at last, that the hydrogen molecule is indeed stable; we had been worried that it might suddenly fall apart!'' Nevertheless, I have always remained particularly proud of having contributed to this result together with J\"urg and his colleagues in Zurich.

Several decades earlier, H$_2$ had already been studied within the framework of the Born-Oppenheimer approximation \cite{1927AnP...389..457B,bransden2003physics}. For a given proton-proton separation $R$%
\footnote{Strictly speaking, one fixes the proton--proton separation rather than the individual proton positions. A more naive presentation of the Born-Oppenheimer method consists of calculating the electronic energy for fixed proton positions, in which case the leading correction arises from the motion of the center of mass.},
the  energy  of the ground state of the two electrons is determined. After adding the direct proton--proton Coulomb repulsion, $1/R$, one obtains an effective potential which is then inserted into the Schr\"odinger equation governing the relative motion of the protons.

This method is remarkably accurate and rests upon profound physical intuition. However, it yields an energy that lies slightly below the exact one and therefore cannot provide a rigorous proof of stability.

More recently, the positronium molecule, Ps$_2$, attracted some attention. In particular, the question arose as to whether it is stable against dissociation into two positronium atoms%
\footnote{Annihilation effects are disregarded and only Coulomb interactions are retained.}. Wheeler appears to have been the first to address this question in a lecture delivered at the New York Academy of Sciences in 1945 and published the following year~\cite{1946NYASA..48..219W}. Using a simple Gaussian trial function%
\footnote{This is somewhat surprising, since Wheeler had already developed sophisticated and powerful methods for solving quantum few-body problems, including the celebrated Resonating Group Method (RGM)~\cite{1937PhRv...52.1107W}.},
he obtained a negative variational energy and thus demonstrated stability with respect to complete dissociation into four isolated particles. However, he was unable to establish that the molecular energy lies below either the positronium-ion threshold or the threshold corresponding to two separated positronium atoms.

In the same year, the Norwegian physicist Ore\footnote{Ore was particularly demanding with respect to proper citation practices. Such rigor would still be welcome today in certain areas of science, where claims of priority and discovery are sometimes rather imaginative; see, for instance, Refs.~\cite{ore1964history,1968Sci...159...56M}.}, then at Yale,
borrowed and adapted techniques previously employed in nuclear physics to study the four-nucleon $\alpha$ particle and concluded that Ps$_2$ was probably unstable~\cite{1946PhRv...70...90O}. Another Norwegian physicist, Hylleraas, already renowned for his pioneering contributions to quantum few-body systems (see, for example, Ref.~\cite{1957qmot.book.....B,alma999221789844702204,alma999221790504702204}), convinced Ore to reconsider the problem using a more rigorous approach. Their subsequent paper, published in 1947, contains a beautiful proof of the stability of Ps$_2$ against dissociation into two positronium atoms~\cite{1947PhRv...71..493H}.

The proof is based on the remarkably ingenious variational ansatz
\begin{equation}
\Psi=
\exp\!\left[-\frac{\alpha}{2}
(r_{12}+r_{24}+r_{14}+r_{23})\right]
\cosh\!\left[
\frac{\beta}{2}
(r_{12}+r_{24}-r_{14}-r_{23})
\right],
\end{equation}
which does not explicitly involve all degrees of freedom but nevertheless possesses sufficient flexibility to produce binding.

It is worth noting that one may set $\alpha=1$ and vary only one parameter. Rather than minimizing directly the variational energy,
$
(\langle T\rangle+\langle V\rangle)/\langle N\rangle$,
one can instead exploit scaling properties and the virial theorem and minimize
$
-\langle V\rangle^2/(
(4\,\langle T\rangle\,\langle N\rangle)$.

Since 1947, the estimate of Hylleraas and Ore has been revisited and progressively improved. For example, Ho employed a trial function consisting of an exponential multiplied by a polynomial in all interparticle distances~\cite{1986PhRvA..33.3584H}. By comparison, the Gaussian-expansion method may appear somewhat heavy, but it ultimately leads to highly accurate energies for the ground state and even predicts an excited state of the positronium molecule~\cite{1998PhRvL..80.1876V}. See also Ref.~\cite{2009NuPhA.827..541C}.

An instructive exercise for students is to rebut the following criticism of Ref.~\cite{1947PhRv...71..493H}:
``... the Hamiltonian of the system was not transformed properly in Ref.~[Hylleraas and Ore, 1947] in order to eliminate the kinetic energy of the center of mass ...''~\cite{1968PhRv..171...36S}.

Surprisingly, the connection between Ps$_2$ and H$_2$ remained largely unexplored for many years. In the early 1970s, papers by Adamowski and collaborators~\cite{1971SSCom...9.2037A,1972PhLA...41..347B}, which we unfortunately discovered only after completing our own work, emphasized the dependence of the binding of systems of the type $M^+M^+m^-m^-$ upon the mass ratio $M/m$. Their analysis was performed in the context of biexcitons in condensed-matter physics. What was missing, however, was a proof that the atom-atom configuration remains the lowest dissociation threshold for arbitrary values of $M/m$, a property that was subsequently demonstrated elegantly by J\"urg and his collaborators in Zurich.

\section{Remarks on Symmetry Breaking}

In almost every introductory course on quantum mechanics, one encounters the elementary exercise of determining the ground-state energy $e(\lambda)$ of the modified one-dimensional harmonic oscillator
\begin{equation}
h(\lambda)=p^2+x^2+\lambda\,x~,
\end{equation}
first perturbatively for small values of $\lambda$ and subsequently by an exact treatment. The exercise is readily generalized by replacing the linear term with an arbitrary odd perturbation, $\lambda\,v_o(x)$. One then finds that
$e(\lambda)\le e(0)$.

In other words, breaking the parity symmetry of $h(0)$ lowers the ground-state energy and therefore strengthens the binding. More generally, the ground-state energy decreases whenever a symmetry-breaking term is added to a symmetric Hamiltonian $H_0$, irrespective of the nature of the symmetry under consideration, be it parity, charge conjugation, isospin, or rotational invariance.

For a few-body system that is stable with respect to its lowest dissociation threshold, however, lowering the total energy does not necessarily imply enhanced stability. Indeed, symmetry breaking often lowers the threshold energy even more significantly, so that the binding relative to the threshold is reduced and may even disappear altogether.

Consider, for instance, the four-body Coulombic system $M^+m^+M^-m^-$. For $M=1$, one recovers the positronium molecule, whose binding energy is known to be very small. In the opposite limit of large $M/m$, the neutral atom $M^+M^-$ becomes extremely compact and is hardly polarized by the light particles. Consequently, the ground state approaches the lowest atom--atom threshold. Detailed studies indeed show that the molecular binding disappears when the mass ratio reaches approximately
${M}/{m}\simeq 2.2$~\cite{1998PhRvA..57.4956B,2005PhR...413....1A}.
Thus, in this example, symmetry breaking is detrimental to stability. We shall see in the next section that breaking charge-conjugation symmetry, starting from the positronium molecule, has precisely the opposite effect and actually enhances stability.
\section{Hydrogen Molecule}
Consider the more general hydrogen-like molecule described by the Hamiltonian
\begin{equation}
H=
\frac{\vec p_1^2}{2\,M}
+\frac{\vec p_2^2}{2\,M}
+\frac{\vec p_3^2}{2\,m}
+\frac{\vec p_4^2}{2\,m}
+\frac{1}{r_{12}}
+\frac{1}{r_{34}}
-\sum_{\genfrac{}{}{0pt}{2}{i=1,2}{j=3,4}}\frac{1}{r_{ij}}~.
\end{equation}
By exploiting the scaling properties of the Coulomb interaction, one may impose
\begin{equation}
\frac{1}{M}+\frac{1}{m}=2~,
\end{equation}
which corresponds to keeping the atom-atom threshold energy fixed at
$E_{\rm th}=-1/2$, and parameterize the inverse masses as
$M^{-1}=1-u$, $m^{-1}=1+u$, $0\le u<1$.
The Hamiltonian can then be rewritten in the remarkably simple form
\begin{equation}
\label{eq:decom:u}
H(u)=H(0)
+\frac{u}{2}
\left(
\vec p_3^{\,2}
+\vec p_4^{\,2}
-\vec p_1^{\,2}
-\vec p_2^{\,2}
\right),
\end{equation}
where the first term is invariant under charge conjugation and the second is odd under this symmetry.
Equation~(\ref{eq:decom:u}) immediately implies that the ground-state energy satisfies
\begin{equation}
E(u)\le E(0)~.
\end{equation}
Hence, a hydrogen-like molecule is necessarily more deeply bound than the Ps$_2$-like system obtained by averaging the inverse masses.
Reference~\cite{1993PhRvL..71.1332R} also contains a proof that the lowest threshold indeed consists of two $M^+m^-$ atoms, together with a variational proof of stability that both improves upon and generalizes the classic treatment of Hylleraas and Ore~\cite{1947PhRv...71..493H}.

% The argument is particularly appealing because it rests on a very general principle. The charge-conjugation-odd component of the Hamiltonian lowers the energy of the four-body system while leaving the atom--atom threshold unchanged. Consequently, the binding relative to the threshold necessarily increases. In this sense, the stability of the hydrogen molecule may be regarded as a direct consequence of the enhancement of binding induced by charge-conjugation symmetry breaking.
%
\section{Stability domain}
\begin{figure}[ht!]
 \centering
 \includegraphics[width=.25\textwidth]{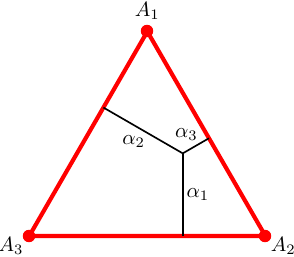}
 \raisebox{-.45cm}{\includegraphics[width=.35\textwidth]{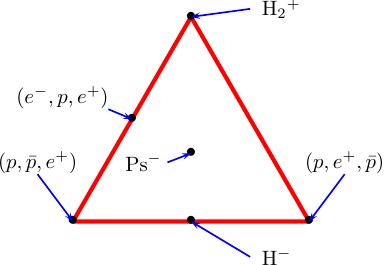}}
\includegraphics[width=.25\textwidth]{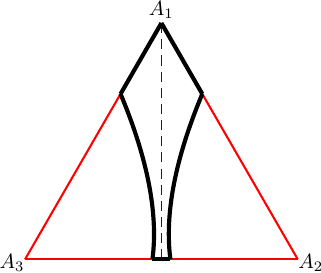}
 % Tri-fig1.pdf: 0x0 px, 300dpi, 0.00x0.00 cm, bb=
 \caption{Stability domain of three unit charges as the function of the inverse masses, normalized to $\sum \alpha_i=1$. Left: definition, center: some configurations, right: stability band. }
 \label{fig:3body}
\end{figure}

For three unit-charge systems, $m_1^\pm m_2^\mp m_3^\mp$, it was found useful to draw the domain of stability inside an equilateral triangle with the inverse masses $\alpha_i=1/m_i$ being the distances to the sides, normalized to $\sum \alpha_i=1$, thanks to the Viviani theorem\footnote{In physics, it is often referred to as the \emph{Dalitz plot} used for three-body decay with the sum of final-state energies being a constant in the center of mass.}.
 See Fig.~\ref{fig:3body}. It can be shown that the stability domain is a band around the symmetry axis $\alpha_2=\alpha_3$, that 1) includes that axis, 2) is convex and 3) is star shaped with respect to the lower vertices.

In the four-body case, the results are less numerous or less rigorous.  Let us mention briefly
\begin{itemize}
 \item Once a certain excess of binding with respect to the threshold is accepted for $M^+M^+m^-m^-$, one can deduce a minimal extension of the stability domain for $M_1^+M_2^+m^-m^-$ with $M_1\neq M_2$ near $M$ or $M^+M^+m_3^-m_4^-$ with $m_3\neq m_4$ near $m$.
 \item A variational ``proof'' of the stability of $m_1^+m_2^+m_3^-m_4^-$ for either $m_1=m_2$ or $m_3=m_4$ has been shown, in which all matrix elements are calculated analytically, but the eigenvalues are estimated numerically~\cite{1997EL.....37..183V}.
 \item Rather intriguing is the existence of Borromean molecules. The word ``Borromean'' was used by nuclear physicists to  designate three-body systems which are bound while all two-body subsets are unbound. The simplest example is the isotope \isotope[6]{He} of Helium considered as a $\alpha n n$ 3-body system with two neutrons and a frozen $\alpha=\isotope[4]{He}$. The isotope \isotope[6]{He}  is stable under strong interactions, while both the di-neutron $nn$ and $\alpha n=\isotope[5]{He}$ are unbound. For more than three bodies, let us define a Borromean state as a system that cannot be built by adding the constituents one by one, resulting into a chain of stable compounds.  Most examples are given with short-range potentials. However, if one considers a molecule $M^+m^+M^-m^-$ with a mass ratio $M/m$ close to 2, one notices that this molecule is stable, but all the three-body subsets are unstable \cite{2003PhRvA..67c4702R}. In the zoo of isotopes and stable hadrons, one can find examples such as $d p \bar d \bar p$ where $p$ is the proton and $d$ the deuteron, or $p K^+ \bar p K^-$ ($K$ designates a Kaon) in the limit of a pure Coulomb interaction.
\end{itemize}

\begin{figure}[ht!]
 \centering
 \includegraphics[width=.5\textwidth]{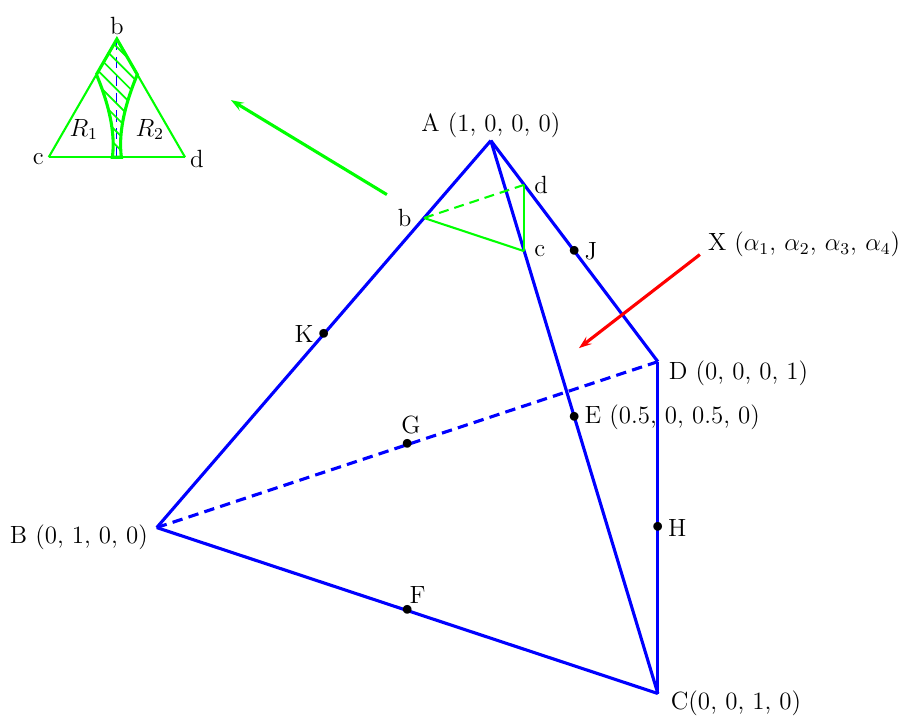}
 % TetraDef.eps: 1804x1433 px, 300dpi, 15.27x12.13 cm, bb=71 356 504 700
 % \includegraphics[width=.3\textwidth]{Figs-Fro-80/Tetrahedron.eps}
\includegraphics[width=.3\textwidth]{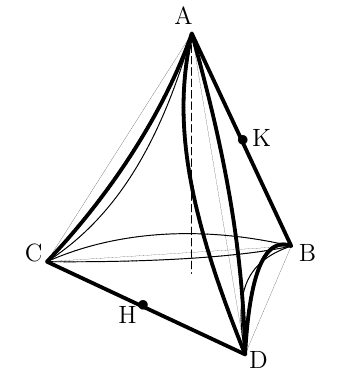}
 % fig-tetra.pdf: 0x0 px, 300dpi, 0.00x0.00 cm, bb=
% TetraDef.eps: 1804x1433 px, 300dpi, 15.27x12.13 cm, bb=71 356 504 700
 \caption{Left: tetraedron of the normalized inverse masses $\alpha_i=1/m_i$ for the $m_1^+m_2^+m_3^-m_4^-$ configurations. Right: sketch of the domain of stability}
 \label{fig:tetra}
\end{figure}

Attempts have been made to sketch the domain of stability in a three-dimensional Viviani plot $\sum m_i^{-1}=1$, inside a regular tetrahedron of unit height. The domain of stability includes the intersections such as $\alpha_1=\alpha_2$ corresponding to $M^+M^+m_3^-m_4^-$ \emph{check}

The overall picture is summarized in Fig.~\ref{fig:tetra}. The notation is such that the vertex $A$ corresponds to the limit in which $m_1$ is much smaller than the remaining masses, whereas the vertex $D$ corresponds to the opposite situation in which $m_4$ is the lightest constituent.

Along the edge $AB$, one has $m_3=m_4$, and the corresponding molecular configurations are therefore stable. Similarly, the edge $CD$, characterized by $m_1=m_2$, belongs entirely to the stability domain.

The midpoint $J$ of the edge $AD$ corresponds to the configuration
$pe^+\bar pe^-$,
in the physical limit $m_p\gg m_e$. This system is unstable and dissociates into protonium and positronium. The same conclusion applies to the midpoint $F$ of the edge $BC$.

By contrast, the points $K$ and $H$, respectively located at the midpoints of $AB$ and $CD$, both correspond to the stable configuration
$
p\,p\,e^-\,e^-$.
Moreover, every point along the bimedian $KH$ represents a stable system of the form
$
M^+M^+m^-m^-$.
The binding relative to the two-atom threshold reaches its minimum at the center of the tetrahedron, which corresponds to the positronium molecule Ps$_2$.

Finally, near the vertex $A$, the system consists of three heavy charges interacting with a light electron. Whenever the heavy three-body ion is stable, the addition of the light particle necessarily produces a stable atom, since the long-range interaction remains attractive and Coulombic. Consequently, as sketched on the right-hand side of Fig.~\ref{fig:tetra}, the four-body stability domain in the vicinity of $\alpha_1=1$ is governed by a replica of the three-body stability band shown in Fig.~\ref{fig:3body}.

\section{Application to the Quark Model}

% This mechanism, although first identified in the context of Coulombic systems, turns out to be considerably more general and will reappear below in connection with exotic hadrons in the quark model.

A closer examination of the argument based on
Eq.~\eqref{eq:decom:u} reveals that its validity depends very little on the Coulombic nature of the interaction. What is essential is, first, the universality of the interaction, namely the fact that particles with different masses experience the same potential, and, second, the existence of a bound state for equal inverse masses lying within the interval bounded by $1/m$ and $1/M$, where $m$ and $M$ denote the light and heavy masses, respectively.

One may therefore formulate a more general statement. Consider four particles with masses
$\{m_i\}=\{M,M,m,m\}$,
interacting through a potential that is independent of the masses and invariant under the simultaneous exchanges
$
1\leftrightarrow 3$,
$2\leftrightarrow 4$.
If a bound state exists for some reference masses $M_0$ and $m_0$, then the system remains stable, and indeed becomes more deeply bound, for
$
M={M_0}/(1-u)$,
$m={m_0}/(1+u)$,
$0\le u<1$.

Provided the the screening parameter $\lambda$ is not too large.
the theorem can be applied, for instance, to screened Coulomb interactions of the form
$1/r/\rightarrow
\exp(-\lambda r)/r$,
which arise in the description of excitons in condensed matter and, more generally, of charges embedded in a polarizable medium.

The quark model was introduced in the early 1960s as a concrete implementation of the rather abstract concepts associated with unitary symmetry, nowadays referred to as flavor SU(3). During the 1970s it evolved into a quantitative framework capable of describing the spectroscopy of charmonium, namely mesons composed of a charm quark $c$ and its antiquark $\bar c$. The construction of potential models became even more demanding when a single interaction was required to account simultaneously for both charmonium and bottomonium, the latter consisting of a bottom quark $b$ and its antiquark $\bar b$
\cite{Quigg:1979vr,Grosse:1997xu}.

In the language of hadron spectroscopy, this universality is known as \emph{flavor independence}. Although flavor independence  receives corrections from spin-dependent terms, it remains one of the fundamental guiding principles of constituent quark models.

For several decades, exotic hadrons have been actively searched for in a wide variety of experiments. By ``exotic'' one usually means a hadronic configuration that cannot be interpreted as either a quark-antiquark pair (an ordinary meson) or a three-quark state (an ordinary baryon). To date, most experimental candidates correspond to resonant states located above their lowest dissociation threshold, and many theoretical treatments still lack the degree of rigor that would be desirable.
% For example, tetraquarks are frequently described as compact bound states of a point-like diquark and a point-like antidiquark, although the dynamical origin of such clustering often remains largely unexplained.

An observation was made in 1981 \cite{Ader:1981db}. Consider a tetraquark of the form $QQ\bar q\bar q$,
with constituent masses arranged as $ MMmm$,
and interacting through a realistic flavor-independent potential.
\begin{enumerate}[i)]
\item
For equal masses, $M=m$,
the tetraquark is unstable against dissociation into two ordinary mesons.%
\footnote{More recently, several studies have identified tetraquark resonances even in this equal-mass case, stimulated in part by intriguing experimental observations at the LHC.}

\item
If charge-conjugation symmetry is broken by introducing unequal masses according to
$
M^{-1}={m_0^{-1}}/(1-u)$,
$m^{-1}={m_0^{-1}}/(1+u)$,
$ 0\le u<1$,
then the tetraquark becomes increasingly cloe to binding as $u$ grows. Above a critical value $u=u_1$, the four-body state eventually becomes stable against strong decay, and when $u$ further increases above $u_1$, the system becomes more and more deeply bound.

\item
For a purely central, spin-independent interaction, the critical mass ratio
${M}/{m} =
(1+u_1)/(1-u_1)$
at which stability first occurs is rather sensitive to the details of the interaction and therefore difficult to predict accurately. In current quark models, this mass ratio turns out to be very large. Fortunately, the physically most interesting states,
$
QQ\bar u\bar d$,
with quantum numbers ($I$ denotes isospin)
$
J^P=1^+$,
$I=0$,
benefit from an additional attractive spin-spin interaction acting within the light antiquark pair. This interaction has no counterpart in the threshold
$
Q\bar u+Q\bar d$,
and therefore contributes directly to the binding of the tetraquark. See, e.g., \cite{Janc:2004qn}.
\end{enumerate}

In 2021, the LHCb Collaboration at CERN reported evidence for an exceptionally narrow state, $T_{cc}^+$,
whose minimal quark content is
$
cc\bar u\bar d$~\cite{LHCb:2021auc}. Its properties turned out to be in remarkable agreement with predictions based on the hydrogen-molecule mechanism discussed above \cite{Ader:1981db}. It means that the enhanced stability induced by breaking charge-conjugation symmetry appears to be realized in Nature.

The corresponding double-beauty state, $T_{bb}=bb\bar u\bar d$,
is expected to lie significantly below its lowest strong-decay threshold and should therefore be a genuinely stable hadron with respect to the strong interaction. Its eventual discovery would open a new chapter in heavy-hadron spectroscopy and in weak-inteaction physics and provide another striking manifestation of the mechanism that underlies the stability of the hydrogen molecule.

\section{Outlook}

A final remark concerns the role of patience in scientific research.
%Patience is perhaps a virtue traditionally associated with the Swiss temperament, and it therefore seems particularly appropriate to emphasize it when celebrating J\"urg's eightieth birthday.
Indeed, the history of few-charge systems and exotic hadrons offers several striking illustrations of the long timescales that may separate theoretical prediction from experimental confirmation. Nearly sixty years elapsed between the proof of the stability of the positronium molecule Ps$_2$~\cite{1947PhRv...71..493H} and its first indirect observation~\cite{2007Natur.449..195C}. Likewise, no less than forty years passed between the prediction of doubly heavy tetraquarks~\cite{Ader:1981db} and the first evidence for the $T_{cc}^+$ state reported by the LHCb Collaboration at the Swiss--French border~\cite{LHCb:2021auc}.

One may reasonably hope that the coming years will bring further developments. On the mathematical side, many aspects of the stability domains of Coulombic few-body systems still await a rigorous understanding. On the phenomenological side, the anticipated discovery of doubly beauty tetraquarks and other multiquark configurations may provide new and unexpected realizations of the same underlying mechanisms. The remarkable interplay between symmetry, stability, and mass asymmetry therefore continues to offer a fertile ground for both mathematical physics and hadron spectroscopy.

\subsection*{Acknowledgments}

It has been both a pleasure and an honour to collaborate with J\"urg on the problem of the stability of the hydrogen molecule, to which Gian Michele Graf and Maximilian Seifert also made essential contributions. I would also like to express my gratitude to my collaborators on various aspects of few-charge systems, namely Edward Armour, Sonia Fleck, Ali Krikeb, Andr\'e Martin, Kalman Varga, and Tai-Tsun Wu.

Finally, I wish to thank Volker Bach, Simeon Reich, and Alexander Zaslavski for initiating this special issue and for providing an opportunity to celebrate J\"urg Fr\"ohlich's scientific achievements and lasting influence on mathematical physics.

%\printbibliography
%	\printbibitembibliography

\end{document}